\documentclass[twocolumn, floatfix,showpacs,preprintnumbers,amsmath,amssymb,pra,superscriptaddress,longbibliography]{revtex4-2}
\usepackage{color}
\usepackage[dvipsnames,svgnames,table]{xcolor}
\usepackage[colorlinks=true,linkcolor=blue,urlcolor=blue,citecolor=blue]{hyperref}
\usepackage{mathtools}
\usepackage{graphicx}
\usepackage{dcolumn}
\usepackage{array}
\usepackage{lipsum}
\usepackage{bm}
\usepackage{subfigure}
\usepackage{amssymb}
\usepackage{multirow}
\usepackage{tabularx}
\usepackage{amsmath}
\usepackage{braket}
\usepackage{csquotes}
\usepackage{enumitem}
\graphicspath{{plots/}}
 \usepackage{lipsum}
\usepackage{mathrsfs}
\usepackage{MnSymbol}

\usepackage{algorithm}
\usepackage{algpseudocode}
\usepackage{todonotes}

\usepackage[normalem]{ulem} 
\newcommand{\xv}{\mathbf{x}}
\newcommand{\vv}{\mathbf{v}}
\newcommand{\kv}{\mathbf{k}}
\newcommand{\pv}{\mathbf{p}}
\newcommand{\qv}{\mathbf{q}}

\newcommand{\rhokkpvec}{\braket{\kv | \rho^{(1)} | \kv'}}

\newcommand{\deltaepsvec}{\varepsilon_\kv - \varepsilon_{\kv'}}

\newcommand{\deltafvec}{f_\kv - f_{\kv'}}
\begin{document}
\title{Mermin's dielectric function and the f-sum rule}

\author{Thomas Chuna}
\email{t.chuna@hzdr.de}
\affiliation{Center for Advanced Systems Understanding (CASUS), D-02826 G\"orlitz, Germany}
\affiliation{Helmholtz-Zentrum Dresden-Rossendorf (HZDR), D-01328 Dresden, Germany}


\author{Jan Vorberger} 
\affiliation{Helmholtz-Zentrum Dresden-Rossendorf (HZDR), D-01328 Dresden, Germany}

\author{Thomas Gawne}
\affiliation{Center for Advanced Systems Understanding (CASUS), D-02826 G\"orlitz, Germany}
\affiliation{Helmholtz-Zentrum Dresden-Rossendorf (HZDR), D-01328 Dresden, Germany}

\author{Tobias Dornheim}
\affiliation{Helmholtz-Zentrum Dresden-Rossendorf (HZDR), D-01328 Dresden, Germany}
\affiliation{Center for Advanced Systems Understanding (CASUS), D-02826 G\"orlitz, Germany}

\author{Michael S. Murillo}%
  \affiliation{Computational Mathematics, Science and Engineering, Michigan State University, East Lansing, Michigan 48824, USA}%
\date{\today}

\begin{abstract}
Mermin’s dielectric function [N.D. Mermin, Phys. Rev. B \textbf{1}, 2362 (1970)] is widely assumed to satisfy the f-sum rule because he constrains his \textit{ansatz} with the continuity equation. However, we identify a moment-closure problem in Mermin's use of the continuity equation. Further, we show that the Mermin's model can be derived without invoking continuity. We describe how other approaches such as the “completed Mermin’’ model of Chuna and Murillo [Phys. Rev. E \textbf{111}, 035206 (2025)] remedy this closure issue. We then inspect the f-sum rule for both the original and completed Mermin models and find for the Mermin \textit{ansatz} that collision frequencies scaling as $\omega$ must violate the f-sum rule, whereas constant, real, positive collision frequencies will satisfy it, with the caveat that, in practice, convergence with respect to the upper integration limit $\omega_{\max}$ is sufficiently slow that finite-domain numerical evaluations exhibit apparent violations, regardless of wavenumber $q$. We also find that collision frequencies with constant imaginary components cause f-sum rule violations. We conclude that if Mermin's model is fit to data via optimizing its collision frequency, then the f-sum rule is not inherently satisfied; constraints, though broad, are needed in order to assume the f-sum rule is satisfied. Further, if the f-sum rule is theoretically satisfied, but violations still appear, then these deviations ought to be included in the error estimates.
\end{abstract}

\maketitle

\section{Introduction}
A variety of experimental efforts make use of dielectric function models for the interpretation of their data. For example, such models are used to analyze x-ray Thompson scattering (XRTS) spectra~\cite{Glenzer_PRL_2007, Plagemann_NewPhys_2012, Sperling_PRL_2015, witte_PRL_2017, fortmann_LasersPartBeam_2009, fortmann_PRE_2010, Graziani2014,Schoerner_PRE_2023},  in solid state optics experiments~\cite{tanuma_electronspec_1993, werner_JPhysChem_2009, shinotsuka_SurfaceInterface_2022, tan_NucInstuments_2025, abril_PRA_1998}, and in stopping power measurements~\cite{CORREA2018291,PhysRevA.58.357,PhysRevE.90.053102,PhysRevE.101.053203}, the latter of which are of key importance, e.g., for inertial fusion energy applications~\cite{vorberger2025roadmapwarmdensematter,drake2018high}.

The quality of a dielectric function is often assessed or improved by its satisfaction of frequency sum rules, which serve as an order-by-order check that important physics considerations are preserved by the model~\cite{selchow_PRE_1999, smith_PRB_1978, Vashishta_PRB-VS-LFC_1972, Singwi_PhysRev-cSTLS1_1968, Sjostrom_PRB-cSTLS2_2013, hasegawa_JPSJ-qSTLS_1975, murillo_HEDP_2008, ailawadi_PRA_1971, arkhipov_PRL_2017, Hansen_PRL-DSF_1974, Hansen_PRA-DSF_1975, Tanaka_PRA_1987, gregori_PoP_2009, arkhipov_CPP_2018, tkachenko_book, Dornheim_moments_2023}. A collection of models, their associated sum rules, and the relevant physics is given by Choi~\textit{et al.}~\cite{Choi_PRE_2019}. Of the frequency sum rules there is one regarded as \textit{the} frequency sum (f-sum) rule, which is the dominant contribution to the intermediate scattering function. Physically, the f-sum rule is understood to indicate whether a model satisfies the continuity equation. 

Across the broader scientific community, there are numerous practical applications of the f-sum rule. Firstly, the f-sum rule is used in XRTS diagnostics to normalize the dynamic structure factor (DSF)~\cite{garcia_Nature_2008, dornheim_SciRep_2024, Dornheim_NatComm_2024, schwalbe2025staticlineardensityresponse, böhme2025correlationfunctionmetrologywarm}. Secondly, the f-sum rule is used in stopping power calculations to normalize the inverse dielectric function~\cite[Section 3.3.2]{hentschel_PhDthesis_2023} or to validate the $\omega$-integration domain~\cite[appendix B]{hentschel_PoP_2023}. Thirdly, Penn's algorithm~\cite{penn_PRB_1987}, which is used in materials science to compute the inelastic mean-free path from experimental optical data~\cite{werner_JPhysChem_2009, shinotsuka_SurfaceInterface_2022, tan_NucInstuments_2025}, requires a dielectric function model that satisfies the f-sum rule. Similar methods are also used in radiation biophysics~\cite{abril_RadiationResearch_2011, nikjoo_ReportsProgressPhys_2016}. As such, the quality of optical data for computing inelastic mean-free path is often assessed using the f-sum rule~\cite{smith_PRB_1978, tanuma_electronspec_1993}.
Essentially, a myriad of fields rely on the assumption that their chosen model of the dielectric function satisfies the f-sum rule. 

Many investigations utilize Mermin's model of the dielectric function under the assumption that it satisfies the f-sum rule because Mermin constrains his \textit{ansatz} via the continuity equation~\cite{mermin_prb_1970}. The authors of Ref.~\cite{Choi_PRE_2019} have assumed this and, within the high energy physics literature, there exist both Mermin-based stopping power calculations~\cite{hentschel_PoP_2023, hentschel_PoP_2025} and Mermin-based XRTS diagnostics~\cite{siegfried_review, Sperling_PRL_2015, github_JaXRTScode}. Additionally, within the solid state literature, there are Mermin-based alternatives to Penn's original algorithm~\cite{abril_PRA_1998, heredia-abril_PRA_2007, denton-abril_physica_2008, da_PRL_2014, nguyen_PhysicalChemC_2015, deVera_PhysChemC_2019}.

The main contributions of this work are as follows. Firstly, we demonstrate that Mermin’s model can be derived without the continuity equation and identify a moment-closure issue that has not, to our knowledge, been discussed. We explain how the completed Mermin (CM) model of Chuna and Murillo resolves this issue~\cite{chuna_PRE_2024}. A key distinction between the models is that Mermin’s model yields a Cauchy distribution in the long-wavelength limit, whereas the CM model yields a Dirac delta distribution. Secondly, based on the Cauchy-like distribution, we find that using the Mermin model with a collision frequency scaling linearly with $\omega$ will violate the f-sum rule, while a constant, real, positive collision frequency will satisfy it. However, even in the latter case, heavy tails impede the numerical evaluation of the f-sum rule, resulting in apparent violations. We discuss prior qualitative reports of convergence issues and provide a numerical investigation of the convergence. Thirdly, we find that introducing a constant imaginary collision frequency, (\textit{i.e.} a frequency shift) causes expected f-sum rule violations in both models.

This publication highlights collision frequency constraints are needed for the Mermin \textit{ansatz} to theoretically satisfy the f-sum rule and highlights the challenges associated with numerically verifying the f-sum rule. Therefore, if Mermin's \textit{ansatz} is assumed to satisfy the f-sum rule, careful consideration must be given to the collision frequency and the size of the integration domain. We encourage f-sum convergence to be considered as an additional source of error as, e.g., done in Ref.~\cite{hentschel_PoP_2025} or when computing the TRK sum rule in DFT Kubo-Greenwood calculations~\cite{calderin_CPC_2017, gawne_PRE_2024}. We also emphasize that when generating synthetic XRTS spectra~\cite{gawne_CPC_2025}, sampling Mermin's \textit{ansatz} will be costly; particularly, if the f-sum rule is to be used to move from arbitrary units to physical units. 

This paper is organized as follows. Section~\ref{sec:Theory} defines sum rules and reviews the connection between the f-sum rule and the continuity equation, identifies the moment closure problem in Mermin's derivation, describes the completed Mermin model that remedies this closure issue, and then analytically evaluates the f-sum rule using the long wavelength limit of both models. Section~\ref{sec:numeric} presents a numeric investigation of the f-sum rule in the quantum and classical cases. Section~\ref{sec:conclusions} presents a summary, draws conclusions for XRTS measurements, and discusses how alternative approaches may address Mermin's moment closure problem.

\section{Theory}\label{sec:Theory}
\subsection{Sum rules of density-density correlators}
The frequency sum rule is a very commonly calculated sum rule and is intimately tied to the continuity equation, but before we investigate this particular sum rule, let us define sum rules for density-density correlators and show why they are important.

The intermediate scattering function $F(q,t)=\braket{n^\dagger(q,t=0) n(q,t)}$ is a two-point correlation function describing the density-density correlation of a system. It is related to the dynamic structure factor by the Fourier transform as 
\begin{align}
    F(q,t) &= \frac{1}{2 \pi} \int^\infty_{-\infty} d\omega\ e^{i \omega t} S(q,\omega) \label{eq:F}
    \\ S(q,\omega) &= \int^\infty_{-\infty} dt\ e^{-i \omega t} F(q,t)\ . \label{eq:DSF}
\end{align}
The intermediate scattering function is used in a host of application and plays a central role in linear response theory~\cite{kubo1966fluctuation, kubo1966fluctuation, boon_molecularhydro_1991, hansen2013theory3rdEd, mahan1990many} and, in many fields, is an important point of connection between theoretical and experimental efforts~\cite{Heaton_PRE_2025, Hao_PRR_2025, Dornheim_NatComm_2024, pratt_PRC_2017}.

In the classical context, the Taylor series expansion of $F$ about $t=0$ is~\cite{boon_molecularhydro_1991, Choi_PRE_2019},
\begin{align}
    F(q,t) &= \omega^{(0)}(q) - \frac{t^2}{2!}\omega^{(2)}(q) + \frac{t^4}{4!}\omega^{(4)}(q) +\ldots \, ,
\end{align}
where the $n$th coefficient is given,
\begin{align}\label{eq:sumrules}
    \omega^{(n)} = \frac{1}{2 \pi} \int^\infty_{-\infty} d\omega \, \omega^n S(q,\omega) \, .
\end{align}
The sum rules are these coefficients \eqref{eq:sumrules} and they determine order by order the correct intermediate scattering function, \textit{i.e.}, the correct structure of two-body correlations. The odd moments are zero because the classical $S(q,\omega)$ is symmetric about $\omega = 0$.

In the quantum setting, the sum rules are given as the coefficients for the Taylor expansion of the response function~\cite{GiulianiVignale_quantumtheory_2008, Dornheim_moments_2023}
\begin{align}
    \text{Re} \, \chi (q,\omega) &= \sum_{n=0} \frac{\omega'^{(n+1)}}{\omega^{n+2}}
\end{align}
where the $(n+1)$th coefficient $\omega'^{(n+1)}$ is given by the sum rules
\begin{align}\label{eq:quantumsumrules}
    \omega'^{(n+1)} = \frac{1}{\pi} \int^\infty_{-\infty} d\omega \, \omega^{n+1} \text{Im} \, \chi(q,\omega).
\end{align}
The even moments are zero due to the odd frequency parity of $\text{Im} \, \chi(q,\omega)$. Notice the quantum version \eqref{eq:quantumsumrules}  recovers the classical version \eqref{eq:sumrules}, (\textit{i.e.}, $\omega'^{(n+1)} \rightarrow \omega^{(n+2)}$) via the classical fluctuation dissipation theorem $S(q,\omega) = 2 \, \text{Im} \, \chi(q,\omega)/\omega$.

\subsection{\textit{The} frequency sum rule and its relation to the continuity equation}
In the classical and quantum literature~\cite{ boon_molecularhydro_1991, GiulianiVignale_quantumtheory_2008} the dominant contribution to the intermediate scattering function (\textit{i.e.}, the lowest order term coefficient in the Taylor expansion) is known as \textit{the} frequency sum (f-sum) rule. \textit{The} f-sum rule is often associated sdawith the continuity equation and here we will demonstrate why. For clarity of presentation, we follow Boon and Yip's classical~\cite{boon_molecularhydro_1991} derivation, but the continuity equation is also important to the quantum version~\cite[eq (3.141)]{GiulianiVignale_quantumtheory_2008}. 

Starting from the definition of the DSF~\eqref{eq:DSF}, we multiply both sides by $\omega^2$ to arrive at~\cite[(2.2.22)]{boon_molecularhydro_1991}
\begin{align}
    \omega^2 S(q,\omega) = \int^\infty_{-\infty} dt e^{-i \omega t} \braket{ \left[ \frac{\partial}{\partial t} n^\dagger(q,t) \right]_{t=0} \left[ \frac{\partial}{\partial t}  n(q,t) \right] } \, .
\end{align}
The two $\partial_t$'s are generated from $\omega^2$ using integration by parts. Next, substituting the continuity equation $\partial_t n(q,t) = \qv \cdot \textbf{j}(q, t) $ produces
\begin{align}
    \omega^2 S(q,\omega) = q^2 \int^\infty_{-\infty} dt e^{-i \omega t}  \braket{ j(q,0) j(q,t)} \, .
\end{align}
This expression along with the assumption that $j(q,t) = \frac{1}{\sqrt{N}} \sum_l \vv_l(t)e^{i q \cdot r_l(t)}$ is used to compute the f-sum rule. The result is~\cite[(2.4.12)]{boon_molecularhydro_1991}
\begin{align}
    \frac{1}{2 \pi}\int^\infty_{-\infty} \frac{d \omega}{\omega_p} \,  \frac{\omega^2}{\omega_p^2} \omega_p S(q,\omega) &= (q/q_D)^2  \, , \label{eq:classicalfsum}
\end{align} 
where we have used the plasma frequency $\omega_p^2 = 4 \pi n_e e^2 / m$, the Debye wavelength $q_D^2= 4 \pi n_e e^2 / T$, and the thermal velocity $v_0^2 = T/m = \omega_p^2/ q_D^2$. 


The f-sum rule has an equivalent formulation, derived in introductory material \cite[Section 3]{tkachenko_book} and \cite[section 4.4]{kremp2005quantum}, 
\begin{align}
    -\pi &= \int_{-\infty}^{\infty} \frac{d\omega}{\omega_p} \,  \frac{\omega}{\omega_p} \, \text{Im} \, \varepsilon^{-1} (q,\omega)\label{eq:fsum_dielectric} \, .
\end{align}
Equation~\eqref{eq:fsum_dielectric} is a general formulation that holds in both the classical and quantum setting and is the formulation we will use to evaluate the f-sum rule. Heuristically, we can arrive at this equation from \eqref{eq:classicalfsum} by substituting the $S(q,\omega)= -\frac{2T}{\omega n} \text{Im} \chi(q,\omega)$ and recognizing the Coulomb interaction $v(q) = \frac{n}{T} (q/q_D)^{-2}$ and the dielectric function $\varepsilon^{-1} (q,\omega) = 1 + v(q) \chi(q,\omega)$.

In summary, if a theory satisfies the f-sum rule then its associated intermediate scattering function is correct at leading order. Additionally, in both quantum and classical systems, the f-sum rule is intimately tied to the continuity equation, as many introductory works invoke continuity to compute the f-sum rule.

\subsection{Deriving Mermin's dielectric function without the continuity equation}\label{sec:flaw}
Often the Mermin ansatz is assumed to satisfy the f-sum rule because it satisfies the continuity equation. However, Atwal and Ashcroft~\cite[eq (12)]{atwal_PRB_2002} derive a generalization of Mermin's model by linearizing the Bhatnagar-Gross-Krook (BGK) kinetic equation~\cite{BGK_PhysRev_1954} via perturbing global equilibrium with respect to local perturbations in chemical potential $\delta \mu$, velocity $\delta \mathbf{u}$, and temperature $\delta T$ and then constraining these perturbations with the lowest collisional invariants (\textit{i.e.}, number, momentum, and energy). Their approach can be greatly simplified and used to derive Mermin's model without invoking the continuity equation, \textit{i.e.}, by neglecting $\delta \mathbf{u}$ and $\delta T$ and constraining $\delta \mu$ via the zeroth collisional invariant. In this subsection, we will do exactly this. \textit{Quantum corrections arise with the inclusion of energy conservation, which is beyond our considerations. So we derive Mermin’s results starting from the classical kinetic equation to provide readers a physical intuition for each term}.

A classical kinetic equation is expressed as
\begin{align}
    (\partial_t + \vv \cdot \nabla_\xv + \textbf{a} \cdot \nabla_\vv) f  = Q \, ,
\end{align}
where the LHS describes the phase-space evolution of the one-body distribution function $f(\xv,\vv,t)$ and the one-body forces are described by the acceleration $\mathbf{a}(\xv,t) = - m^{-1} \nabla_\xv U(\xv,t)$, as such $(\vv \cdot \nabla_\xv + \textbf{a} \cdot \nabla_\vv) f$ is equivalent to a Poisson bracket. The RHS collects all the two-body terms into the collision operator $Q$. 

Following Mermin, we assume the BGK relaxation time approximation (RTA) of the collision term. Thus, the RHS is expressed as the difference between the current distribution and local equilibrium $\mathcal{M}$,
\begin{align}\label{eq:RTA}
    (\partial_t + \vv \cdot \nabla_\xv +\textbf{a} \cdot \nabla_\vv) f  = \nu (\mathcal{M} - f) \, .
\end{align}
Here the collision frequency $\nu = 1/\tau$ is the inverse of the relaxation time $\tau$. Though Mermin does not justify his use of the RTA, the equation has been derived directly from the Boltzmann equation~\cite{haack_JSP_2017, garzo_PhysFluidsA_1989} and the applicability of the model to linear response has been discussed~\cite{chuna_PRE_2024}. 

To produce Mermin's dielectric function, we first focus on the LHS. We expand the one-body distribution function $f$ about global equilibrium as $f(\xv, \vv, t) = f_0(v) + \delta f(\xv, \vv, t)$, Fourier transform, and convert to discrete quantum states  ($f_0(v) \rightarrow f_k \delta_{\kv\kv'}$, $\delta f(\xv, \vv, t) \rightarrow \delta f_{\kv\kv'}$ $\partial_t \rightarrow - i \omega$, $\partial_\xv \rightarrow i \qv$, $\pv = m \vv$, $\kv=\pv+\qv/2$, $\kv'=\pv-\qv/2$) as 
\begin{subequations}\label{eq:FourierTransform}
\begin{align}
    \mathcal{F} \{ \partial_t (f_0 + \delta f) \} &= -i \omega \, \delta f_{\kv\kv'} \, ,
    \\ \mathcal{F} \{ \vv \cdot \nabla_\xv (f_0 + \delta f)\} &=i \frac{\pv}{m} \cdot \qv \, \delta f_{\kv\kv'}  \nonumber
    \\ & =i( \deltaepsvec) \delta f_{\kv\kv'} \, ,
    \\ \mathcal{F} \{ \textbf{a} \cdot \nabla_\vv f_0 \} & = -i\qv U_\mathrm{ext} \cdot \nabla_\pv f_\kv \delta_{\kv\kv'} \nonumber
    \\ &= -i (\deltafvec ) U_\mathrm{ext} \, . 
\end{align}
\end{subequations}
The second equalities follow from Taylor expanding about $\pv$. 
Inserting \eqref{eq:FourierTransform} into the LHS of \eqref{eq:RTA} we have
\begin{align}\label{eq:fourierkinetic}
    \omega \delta f_{\kv\kv'} - (\deltaepsvec) \delta f_{\kv\kv'}  + (\deltafvec) U = i \nu \left( \mathcal{M} - f\right) \, .
\end{align}

We focus the RHS and expand the relaxation term about global equilibrium using the relations
\begin{subequations} \label{eq:localequilibrium}
\begin{align}
    f &= f_\kv \delta_{\kv\kv'} + \delta f_{\kv\kv'},
    \\ \mathcal{M} &\approx f_\kv \delta_{\kv\kv'} - \frac{\deltafvec}{\deltaepsvec}\delta \mu (\kv-\kv')\ .\label{eq:MerminExpansion}
\end{align}
\end{subequations}
Following Mermin, the expanded local equilibrium $\mathcal{M}$ \eqref{eq:MerminExpansion} contains linear deviations from the global equilibrium with respect to the chemical potential $\delta \mu$. Inserting these expansions into the RHS of \eqref{eq:fourierkinetic} produces the linearized kinetic equation,
\begin{align}\label{eq:merminkinetic}
    &\omega \delta f_{\kv\kv'} - (\deltaepsvec) \delta f_{\kv\kv'} + (\deltafvec ) U(\kv-\kv')  \nonumber
    \\ & =- i \nu\left( \delta f_{\kv\kv'} + \frac{\deltafvec}{\deltaepsvec}\delta \mu(\kv-\kv')\right).
\end{align}
Equation \eqref{eq:merminkinetic} matches Mermin~\cite[eq (4)]{mermin_prb_1970}, where Mermin distributed the negative to have $(f_{\kv'} - f_\kv ) U(\kv-\kv')$ and denotes our $\delta f_{\kv\kv'}$ as $\rhokkpvec$.

Equation \eqref{eq:merminkinetic} is not closed, \textit{i.e.}, the $\delta \mu$ in \eqref{eq:merminkinetic} is yet unknown. To constrain this term, Mermin converts $\kv$'s to $\pv$'s using the previous $\kv=\pv+\qv/2$, $\kv' = \pv - \qv/2$ relation (\textit{i.e.}, $\deltaepsvec \rightarrow (\pv\cdot \qv) / m$) and then integrates \eqref{eq:merminkinetic} over momentum $p$, which removes the contribution from the potential term $U$ and introduces $\delta n(\qv,\omega) = \int d \pv \, \delta f_{\kv\kv'}$ (units of number density),
\begin{align}\label{eq:fourierkinetic-linearorder}
%
    &\omega \delta n - \int \frac{d\pv}{4\pi^3} \frac{\pv\cdot \qv}{m} \delta f_{\kv\kv'}\nonumber
    \\ &= - i \nu \left(\delta n + \delta \mu \int \frac{d \pv}{4\pi^3} \frac{ \deltafvec}{\deltaepsvec} \right).
\end{align}
We recognize the second term on the RHS as the quantum ideal gas susceptibility at $\omega=0$, often denoted $\chi_0(\qv,\omega=0)$ and defined in Mermin's paper~\cite{mermin_prb_1970} as, 
\begin{align}\label{eq:quantum_susceptibility}
    \chi_0(\qv,\omega) = \int \frac{d \pv}{4\pi^3} \frac{ \deltafvec}{\omega - (\deltaepsvec)} \, ,
\end{align}
Mermin refers to the LHS of \eqref{eq:fourierkinetic-linearorder} as ``$\omega \, \delta n = \qv \cdot \delta \textbf{j}$'', so that solving for $\delta \mu$ such that the RHS is zero [\textit{i.e.}, constraining $\delta \mu$ to $\delta \mu = \delta n \, / \, \chi(\qv,0)$] enforces the continuity equation.  

However, there is a moment closure problem in Mermin's invocation of the continuity equation. This can be understood by considering the current's dimensions. According to the product rule, a current perturbation $\delta \mathbf{j}$ goes as, 
\begin{align}\label{eq:actualcurrent}
    \delta \mathbf{j}= \mathbf{u}_0 \delta n  + n_0 \delta \mathbf{u},
\end{align}
where $n_0$ and $\mathbf{u}_0$ are the equilibrium mass density and velocity. Yet, local perturbations with respect to velocity $\delta \mathbf{u}$ are not included in the expansion about global equilibrium \eqref{eq:localequilibrium}. 

Let us revisit the constraint that the RHS of \eqref{eq:fourierkinetic-linearorder} is zero, this can be expressed as 
\begin{align}\label{eq:collisioninvariant}
    0= \int \frac{d\pv}{4\pi^3} \left( \delta f_{\kv\kv'} + \frac{\deltafvec}{\deltaepsvec}\delta \mu(\kv-\kv') \right).
\end{align}
Since all the terms in the integrand arise from the collision integral, then \eqref{eq:collisioninvariant} is precisely $\int d \pv \, Q = 0$ at linear order. Therefore, the Mermin \textit{ansatz} is better understood as enforcing the zeroth collisional invariant, (\textit{i.e.} local number conservation), and not as enforcing the continuity equation.

\subsection{Dynamic response models that satisfy the continuity equation}\label{sec:remedy}
A remedy for this moment closure problem is extending the initial expansion about global equilibrium \eqref{eq:localequilibrium} to include local perturbations with respect to velocity $\delta \textbf{u}$. Fortunately, this has already been investigated~\cite{atwal_PRB_2002, chuna_PRE_2024}. 

In both of these prior investigations, $\delta \textbf{u}$ is constrained by momentum conservation (\textit{i.e.} the first collisional invariant) and enforces that~\cite[eq (12)]{atwal_PRB_2002},
\begin{align}\label{eq:momentum_conservation}
    \qv \cdot \delta \textbf{j} =  n_0 \, \qv \cdot \delta \textbf{u} \, ,
\end{align}
By comparison to \eqref{eq:actualcurrent} this implies that $\mathbf{u}_0=0$. Since the moment closure issue is resolved, we can invoke the continuity equation $\qv \cdot \delta \textbf{j} = \, \omega \, \delta n$ to express $\delta u$ in terms of $\delta n$,
\begin{align}
    \delta \textbf{u} =  \frac{\omega}{q^2 n_0} \qv \, \delta n \, . 
\end{align}
This is an additional number density perturbation arising from local velocity deviations and it manifests in the density-density response function as~\cite{chuna_PRE_2024} 
\begin{align} \label{eq:singlespeciesCM}
    \chi^\textrm{CM} = \cfrac{C_0}{ 1 - \cfrac{i\nu}{\omega + i \nu} \left( 1 - \cfrac{C_0}{B_0} \right)  - i \nu \frac{m \omega}{ q^2 n_0}C_0 }.
\end{align}
where 
\begin{subequations}
\label{eq:classical_susceptibility}
\begin{align}
    C_0(\qv,\omega) &\equiv \int d \vv \frac{m_i^{-1} \qv \cdot \nabla_\vv f(v)}{\vv \cdot \qv - \omega}, \label{eq_Cn}
    \\ B_0 &\equiv  C_0(\qv,0). 
\end{align}
\end{subequations}
and $\pv = m \vv$. We have suppressed the $\qv$ and $\omega$ dependence in $C_0$ and $B_0$ for presentation. Equation \eqref{eq:singlespeciesCM} is known as the completed Mermin (CM) susceptibility $\chi^\textrm{CM}(\qv, \omega + i \nu)$ without mean field correction; the full derivation can be found in Ref.~\cite{chuna_PRE_2024}. For comparison, we present the single species Mermin susceptibility $\chi^\textrm{Mermin}(\qv, \omega + i \nu)$, also without mean field correction,~\cite{chuna_PRE_2024}
\begin{align} \label{eq:singlespeciesM}
    \chi^\textrm{Mermin} = \cfrac{C_0}{ 1 - \cfrac{i \nu}{\omega + i \nu} \left( 1 - \cfrac{C_0}{B_0} \right) }.
\end{align}
Notice this expression lacks the term arising from $\delta \mathbf{u}$ in the denominator.

For the quantum version of \eqref{eq:singlespeciesCM} and \eqref{eq:singlespeciesM}, simply replace $C_0$ and $B_0$ with the corresponding quantum ideal gas susceptibility. Additional alterations beyond this swap, are only needed if $\delta T$ is included and the energy conservation is enforced~\cite{atwal_PRB_2002}. Our numeric implementations of these response functions in the quantum and classical case are available on GitHub~\cite{github_DynamicResponsecode}. Our implementation of the quantum case follows Giuliani and Vignale~\cite{GiulianiVignale_quantumtheory_2008} and Tolias \textit{et al}.~\cite{tolias_CPP_2024}, while our implementation of the classical case follows Ichimaru~\cite[Ch 4.]{ichimaru2018plasmavol1}.


\subsection{Analytic form of the inverse dielectric function in the long wavelength limit}
Now that we have discussed the moment closure issue in the continuity equation, we turn our attention to whether the f-sum rule is satisfied. In this section we will evaluate the long wavelength limit analytically. We consider the homogeneous one-component plasma (OCP), with its Coulomb interaction potential $v(q) = \frac{4 \pi e^2}{q^2} =  \frac{T}{n} \frac{q_D^2}{q^2}$. The long wavelength limit of the CM model for real $\tau$ at order $\mathcal{O}[q^2]$ is given by~\cite{chuna_PRE_2024}
\begin{align}\label{eq:CMlongwavelength}
    \underset{q \rightarrow 0}{\lim} \: \text{Im}\left\{\frac{1}{\varepsilon^\textrm{CM}} \right\} &=- \frac{ q^2 \cfrac{ \omega_p^4 \, \omega/ \nu}{1 + (\omega/\nu)^2}  }{ q_D^2 (\omega^2 - \omega_p^2)^2  + 2 q^2 \cfrac{ \omega_p^2 (\omega_p^2 - \omega^2)}{1 + (\omega/\nu)^2}} \nonumber
    \\ &= - \pi \delta (\omega - \omega_p ) \, ,
\end{align}
The CM model \eqref{eq:CMlongwavelength} is a Cauchy-like distribution of vanishing width $q^2$, \textit{i.e.}, a Dirac delta function. Inserting a Dirac delta function into \eqref{eq:fsum_dielectric} shows that the f-sum rule is trivially satisfied. 

By comparison, Atwal and Ashcroft~\cite[eq (36)]{atwal_PRB_2002} have demonstrated that Mermin's dielectric function does not converge to a Dirac delta function in the long wavelength limit. This directly disagreed with an earlier publication from Selchow~\textit{et al.}~\cite{selchow_PRE_1999}, who later published an erratum~\cite{selchow_PRE-erratum_2004} in agreement. Instead of a Dirac delta, Mermin's inverse dielectric function \eqref{eq:singlespeciesM} recovers a finite width Cauchy-like distribution~\cite{atwal_PRB_2002, chuna_PRE_2024, Svensson_PRE_2025}
\begin{align}\label{eq:Merminlongwavelength}
    &\underset{q \rightarrow 0}{\lim} \: \text{Im}\left\{\frac{1}{\varepsilon^\textrm{Mermin}} \right\} = - \frac{  \nu \, \omega \, \omega_p^2}{ ( \omega^2 - \omega_p^2)^2 + \nu^2 \omega^2}.
\end{align}
Such Cauchy/Lorentzian distributions occur commonly when computing hydrodynamic dynamic structure factors, where the sum rules are known to diverge; see Hansen and McDonald~\cite[Ch 8.6]{hansen2013theory3rdEd} for a thorough treatment (Footnote \footnote{Essentially, the issue arises because the distribution decays too slowly to define the integral from $\pm \infty$. This is discussed in introductory statistics textbooks~\cite[section 7.1.2]{blitzstein2019introduction} and follows readily from the introductory calculus example known as ``Gabriel's horn'' where students show integral of $1/x$ diverges, but the integral of $1/x^2$ does not. Essentially, multiplying the Cauchy distribution by $x$ causes it to scale like $1/x$ at large $x$ and thus its integral diverges.}).

Though Mermin's model \eqref{eq:Merminlongwavelength} resembles a Cauchy distribution, it has an $\omega$ dependent decay width, so the function scales as $\nu \, \omega^{-3}$ at large $\omega$ (this matches Morawetz and Fuhrmann~\cite{Morawetz_PRE-momentum_2000}). The f-sum rule introduces an additional factor of $\omega$, causing the integrand to scale as $\nu \, \omega^{-2}$ at large $\omega$. Therefore, the f-sum rule is convergent. However, if there is a linear $\omega$ dependence in the collision frequency, (\textit{i.e.}, $\nu(\omega) \sim \omega$), then the integral will diverge. 

\subsection{Analytic evaluation of the f-sum rule in the long wavelength limit}
To better understand the convergence, we analytically evaluate the f-sum rule~\eqref{eq:fsum_dielectric} using Mermin's long wavelength limit \eqref{eq:Merminlongwavelength} with a constant, positive, real collision frequency. We find that the integral evaluates to $-\pi$ as desired
\begin{align}\label{eq:Merminfsumconvergence}
    - \pi = \lim_{\omega_{\max} \rightarrow \infty}  \left\{ -2 \arctan\left(\frac{\omega_{\max}^2- \omega_p^2}{\omega_{\max} \, \nu}\right) \right\} \, .
\end{align}
Here $\omega_{max}$ quantifies the upper and lower integration bounds normalized by the plasma frequency $\omega_p$; for the derivation see Appendix~\ref{app:complexanalysis}. As a sanity check, notice that in the Lindhard limit, $\nu \rightarrow 0$, the f-sum rule is satisfied for any $\omega_{\max} > \omega_p$, a clear sign of a Dirac delta peak. 

Equation \eqref{eq:Merminfsumconvergence} establishes that, in the long wavelength limit with real positive collision frequency, Mermin's dielectric function   satisfies the f-sum rule. However, the convergence of $\arctan(\omega_{\max})$ to $\pi/2$ is slow in a numerical sense. So while theoretically the bounds can easily be taken to $\pm \infty$, in practice finite numeric bounds lead to apparent violations of the f-sum rule.

\section{Numeric Results}\label{sec:numeric}
\subsection{Evaluations of the f-sum rule for real collision frequency}
We indicate previous investigations that evaluated the f-sum rule as part of their publications; each of which qualitatively observed f-sum convergence issues~\cite{Morawetz_PRE-XC_2000, arkhipov_NASRK_2013, da_SurfaceInterface_2019, roepke_PRE_1998, chuna_PRE_2024}. In particular, we quote Morawetz and Fuhrmann~\cite{Morawetz_PRE-XC_2000} who comment that ``[the f-sum rule] is fulfilled numerically for all approximations, however, the convergence is very bad for [Mermin's model] or [Mermin's model with an additional energy conservation correction]. The inclusion of momentum conservation in turn improves the convergence of the sum rule appreciably.''. Other than describing this phenomena, further investigation is not given. In this section, we give insight into these convergence issues, providing a quantitative characterization.

We investigate the homogeneous electron gas at $r_s = 2.07$, $\Theta=0.0855$, and $q= 0.88 q_F$, with $\nu = \tau^{-1} = 0.5 \, \omega_p$. These conditions and associated collision frequency are reflective of the large $\omega$ behavior for Aluminum at 1 eV presented by Hentschel \textit{et al}.~\cite[Figure 4]{hentschel_PoP_2023}; specifically for the transfer matrix (TM) cross section with inelastic contribution (inel) and the Kubo-Greenwood (KG) collision frequency. We use both quantum (Fermi-Dirac statistics) and classical (Maxwellian statistics) ideal gas susceptibilities. We compute the f-sum rule for $\chi^{CM}(q,\omega + i \nu)$ and $\chi^{Mermin}(q,\omega + i \nu)$ at a real collision frequency. We evaluate the f-sum rule using a discrete sum, where the upper and lower bounds are $\pm \omega_{\max} = \pm N_{\omega} \Delta \omega$ and $\Delta \omega$ is the discretization width. The numerical expression is given
\begin{align}
    \sum_{n =-N_\omega}^{+N_\omega} \frac{\Delta \omega}{\omega_p} \,  \frac{\omega_n}{\omega_p} \, \text{Im} \, \varepsilon^{-1} (q,\omega_n + i \nu)  \, , \label{eq:fsum_numeric}
\end{align}
here $\omega_n = n  \Delta\omega$. 

\begin{figure}
    \centering
    \includegraphics[width=\linewidth]{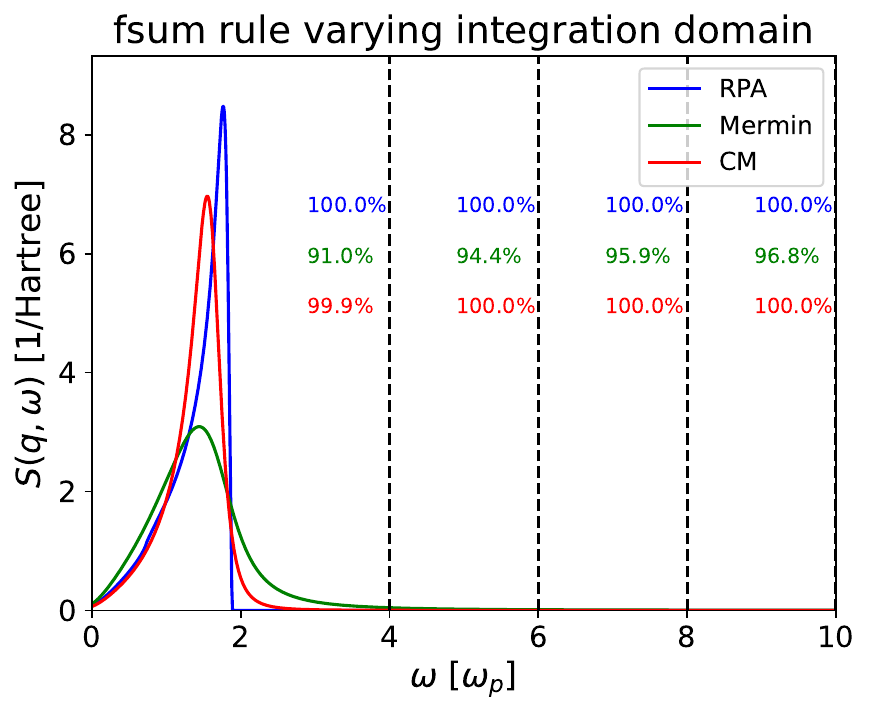}%
    \caption{ Plot of the DSF obtained at wavenumber $q=0.88 q_F$, $\Theta=0.0855$, $r_s=2.07$, and $\nu = 0.5 \, \omega_p$ for the quantum Fermi-Dirac distribution; these condition are reflective of the large $\omega$ limit for Aluminum at 1 eV, see Hentschel~\textit{et al.}~\cite{hentschel_PoP_2023}. The percentage to which the f-sum rule is satisfied at a given $\omega_{\max}$ is indicated by colored text adjacent to the vertical line. The color and order matches that of the legend.}
    \label{fig:fsum_omegadomain}
\end{figure}

Firstly, we plot the DSF at fixed $q=0.88 q_F$ in Figure~\ref{fig:fsum_omegadomain}. Dashed vertical lines indicate where the integration bound must be to achieve a certain f-sum rule convergence. All curves are indistinguishable from 0 for $\omega / \omega_p \in [4,10]$, yet Mermin's model still has an f-sum rule violation of $\approx 3\%$ at $\omega = 10  \, \omega_p$. This plot demonstrates that the integration domain must greatly exceed the domain where Mermin produces interesting behavior in order to have f-sum rule convergence. 

\begin{figure}[h!]
    \centering
    \includegraphics[width=\linewidth]{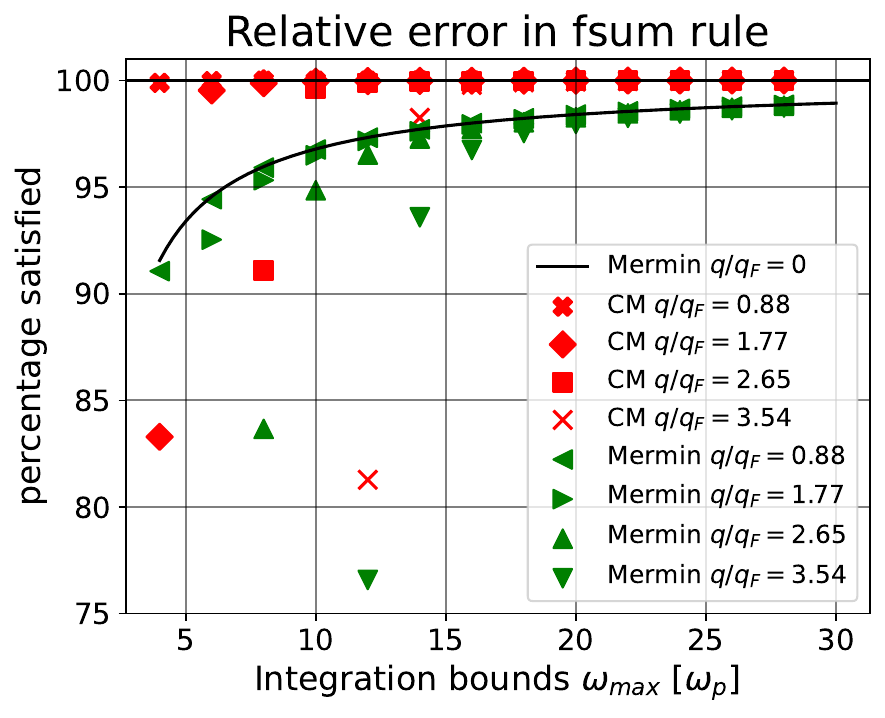}
    \caption{Plot of the relative error in the frequency sum (f-sum) rule~\eqref{eq:fsum_dielectric} of the Mermin (green) and completed Mermin (CM, red) models computed using \eqref{eq:fsum_numeric} with grid spacing $\Delta \omega = \omega_p/100$ and variable integration bounds. These computations are carried out at $r_s = 2.07$, $\Theta=0.0855$, $\nu = 0.5 \, \omega_p$, and four different wavenumbers $q/q_F$. For comparison, we plot the long wavelength limit of Mermin's f-sum rule convergence \eqref{eq:Merminfsumconvergence} as a black line.}
    \label{fig:fsum_convergence}
\end{figure}

Secondly, we compute the relative deviation from the expected f-sum rule value as a function of the integration bounds at various wavenumbers. We plot these numeric results alongside the analytic long wavelength expression \eqref{eq:Merminfsumconvergence}, see Figure~\ref{fig:fsum_convergence}. The finite $q$ values match the analytic $q \rightarrow 0$, scaling as $\arctan$ at large $\omega_{\max}$, thus demonstrating the validity of the analytic expression.

\begin{figure}[h!]
    \centering
    \includegraphics[width=\linewidth]{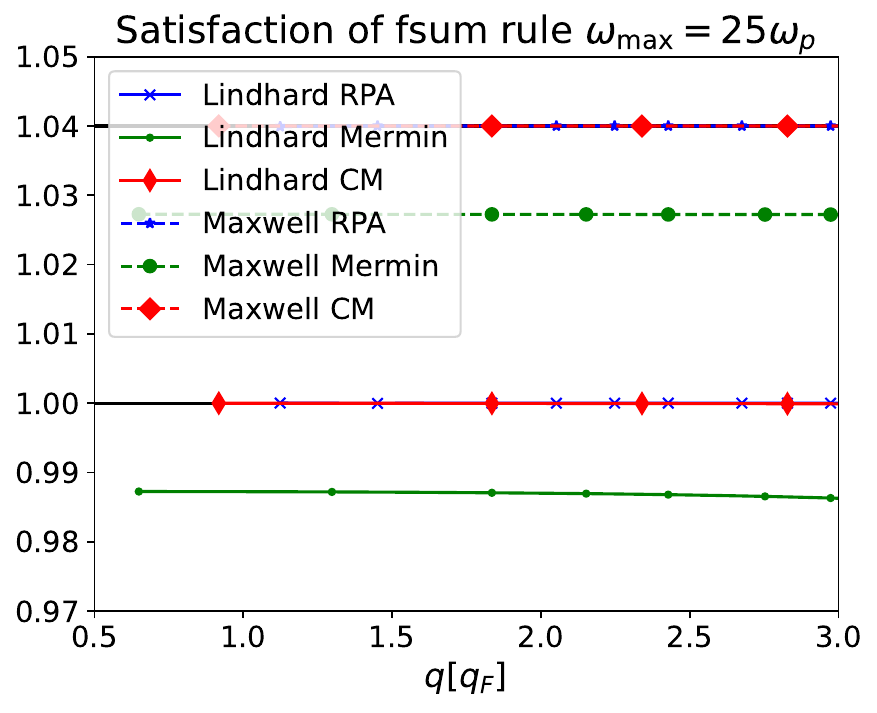}%
    \caption{Plot of the ratio of the inverse dielectric function's first moment, evaluated via \eqref{eq:fsum_numeric}, to the expected value of $-\pi$ for the RPA (blue), Mermin (green) and completed Mermin (CM, red) models across wavenumber $q$ at $r_s = 2.07$, $\Theta=0.0855$, and $\nu= 0.5 \, \omega_p$ using both the classical Maxwellian equilibrium distribution and the quantum Fermi-Dirac distribution. Notice that the classical computations have been displaced by $\delta=0.04$, so the relative error is equivalent to the Lindhard case.}
    \label{fig:fsum_epsinv}
\end{figure}

Lastly, we gain additional insight into how the f-sum convergence depends on $q$ and on Maxwellian statistics. Though the assumption of Maxwellian statistics is not physical, replacing the Fermi ideal gas susceptibility with a classical susceptibility provides insight about the functional form of Mermin's \textit{ansatz}. We evaluate \eqref{eq:fsum_numeric} with $\Delta \omega = \omega_p/100$ and integration bounds $\pm \omega_{\max} = \pm 25$ and present the results in Figure~\ref{fig:fsum_epsinv}. We find the deviation from the expected value does not depend strongly on any of these choices. 

\subsection{Evaluation of the f-sum rule for constant imaginary frequency}
\begin{figure}[h!]
    \centering
    \includegraphics[width=\linewidth]{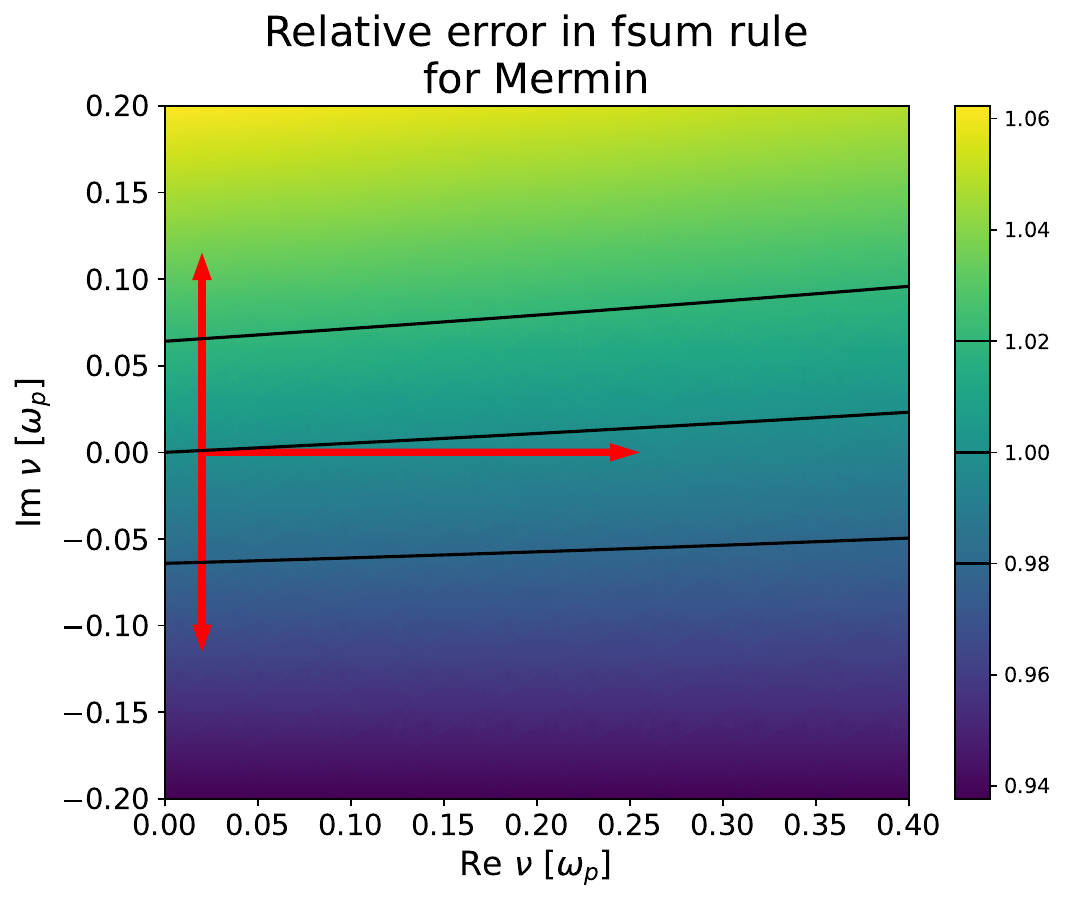}
    \includegraphics[width=\linewidth]{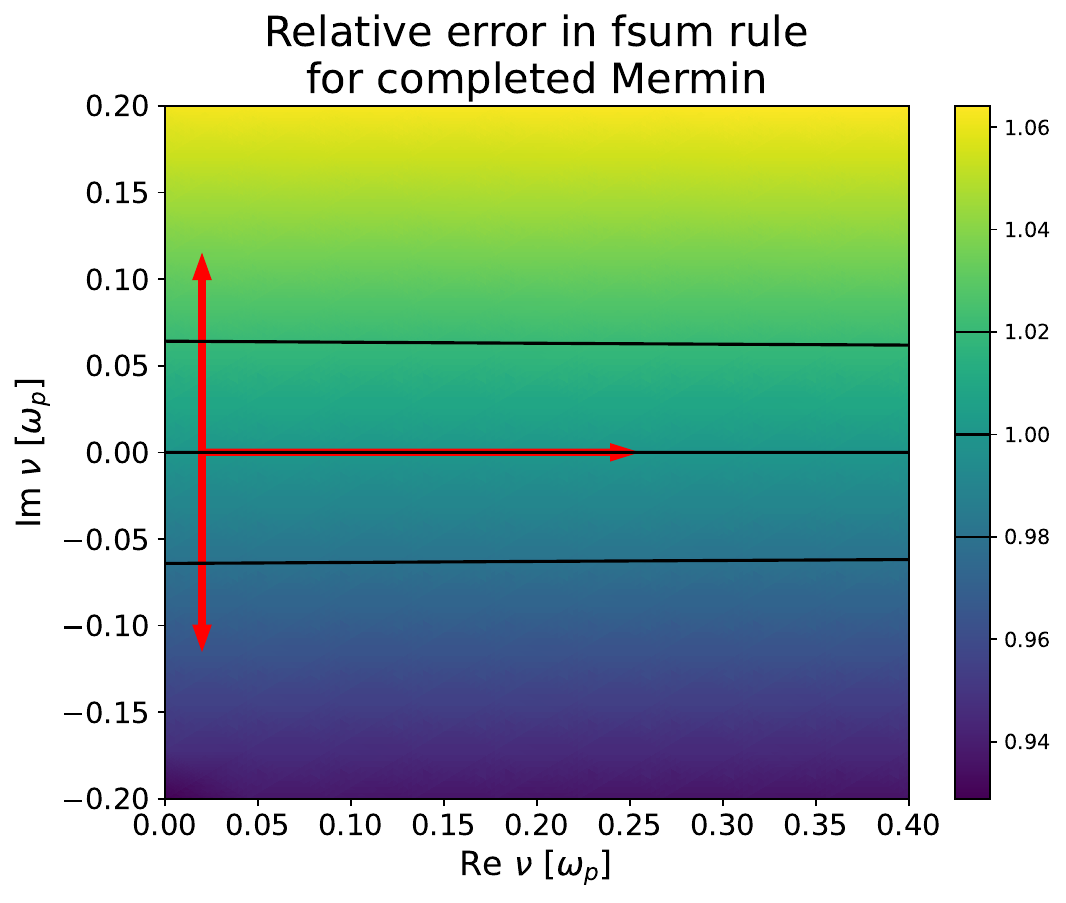}
    \caption{Heatmaps indicating the ratio of the inverse dielectric function's first moment, computed via \eqref{eq:fsum_numeric}, to the expected value of $-\pi$. Evaluations conducted at $r_s = 2.07$, $\Theta=0.0855$, $q=0.88 q_F$, and $\omega_{\max} = 40 \, \omega_p$ with Fermi-Dirac statistics with different constant collision frequencies $\nu(\omega) = \nu_0$, where $\nu(-\omega) = \nu_0^*$ so that its Fourier transform is real. Red arrows start from the RPA approximation $\nu = 0 + i 0$ and move outward. The rightward arrow indicates the effect of increasing the peak width. The upward arrow indicates the effect of shifting the peak to larger $\omega$, and the leftward arrow indicates the effect of shifting the peak to smaller frequencies. Notice the color bars differ.}
    \label{fig:fsum_heatmap}
\end{figure}

While the collision frequency is expected to be only real in the large $\omega$ and small $\omega$ limits, it is still interesting to consider the effect of a constant imaginary collision frequency; this will demonstrate the effect of shifting the peak location of the DSF. We provide a heatmap in Figure~\ref{fig:fsum_heatmap} that indicates the first moment (on a fixed integration interval $[-40 \omega_p, \, +40 \omega_p]$) relative to to the expected value $-\pi$. 

We observe from the rightward arrow in Figure~\ref{fig:fsum_heatmap} that for increasing real collision frequency, more of the Mermin model's weight is shifted outside of the integration interval, leading to a diagonal slope in the contours that is not present in the completed Mermin model. This effect persists regardless of the complex magnitude (\textit{i.e.}, shift in the plasmon peak). We observe from the upward and downward arrows that shifting the peak to larger and smaller frequencies (\textit{i.e.}, blue and red shifts) causes the first moment to differ from $-\pi$. This is to be expected, since shifts in the distribution's centroid alter which frequencies are more or less weighted. 

\section{Summary, Conclusions, and Future Work} \label{sec:conclusions}
In the first half of this paper we have shown that Mermin's derivation has a moment closure problem and that the remedy is extending Mermin's expansion about local equilibrium to include not just deviations from the local chemical potential, but also local velocity. As such, Mermin's \textit{ansatz} is best understood as enforcing the zeroth collisional invariant (\textit{i.e.} local number conservation) and not the continuity equation. In this work, we discuss two prior works~\cite{atwal_PRB_2002, chuna_PRE_2024} that constrained the local velocity perturbation using the continuity equation and the first collisional invariant (\textit{i.e.}, conservation of momentum), leading to the so-called ``completed Mermin'' (CM) dielectric function. 

In the second half of this paper and due to the connection between the f-sum rule and the continuity equation, we consider carefully whether Mermin's \textit{ansatz} satisfies the f-sum rule. We find for collision frequencies  that scale as $\omega$ Mermin's first moment does not converge to the desired f-sum rule value. For constant positive collision frequencies, Mermin's first moment converges to the desired f-sum rule value, as characterized by \eqref{eq:Merminfsumconvergence}. The severity of this $\arctan$ convergence has been made clear by Figure~\ref{fig:fsum_omegadomain}, Figure~\ref{fig:fsum_convergence}, and Figure~\ref{fig:fsum_epsinv}. In practice any reasonably finite $\omega$ grid can incur a non-negligible error that ought to be accounted for in the error analysis; these are the consequences of Mermin's Cauchy-like distributions. Finally, we demonstrate that a constant shift in the peak position manifests as a violation of the f-sum rule, see Figure~\ref{fig:fsum_heatmap}. This is expected since it is by definition a statement about the first frequency moment.

We conclude the following for the traditional approach to extracting collision frequencies from XRTS, \textit{i.e.}, fitting a $q$- and $\omega$-dependent collision frequency to some observed signals~\cite{Sperling_PRL_2015, witte_PRL_2017, Schoerner_PRE_2023, hentschel_PoP_2025}. The $q$- and $\omega$-dependent collision introduces significant flexibility, allowing the Mermin \textit{ansatz} to reproduce a broad class of functional forms. However, unconstrained fits may violate the physical constraints on the collision frequency, which are: firstly, where the Mermin \textit{ansatz} is expected to recover the f-sum rule, the imaginary part of the collision frequencies must go to zero in the long wavelength limit; secondly, where the collision frequency $\nu(\omega)$ is interpreted as an impurity scattering frequency, then for small $\omega$ the imaginary part of the collision frequency must scale linearly with $\omega$, \textit{i.e.}, $\text{Im} \, \nu(\omega) \sim \omega$, going to zero for $\omega=0$. See \cite[Section 4.6.1]{GiulianiVignale_quantumtheory_2008} as well as \cite{Reinholz_PRE_2000} for other constraints. Such constraints were not incorporated in some earlier XRTS analyses. See, e.g., Ref.~\cite[supplemental material]{Sperling_PRL_2015}, where the imaginary component grows larger with decreasing $\omega$.  We encourage future work to explore the line of research pursued by Hentschel \textit{et al}.~\cite{hentschel_PoP_2025} who derived a new sum rule and enforced it on their collision frequency fits.

There are additional conclusions for XRTS ray tracing codes, which use Monte Carlo methods to generate synthetic experimental spectra from a DSF model and normalize over counts to match the f-sum. Figure~\ref{fig:fsum_omegadomain} indicates that if the Mermin \textit{ansatz} is used, the synthetic XRTS spectra may need to sample an unphysically large spectral range to achieve the f-sum rule convergence. These convergence issues are expected to improve for frequency-dependent collision frequencies that go to zero at large $\omega$ such as Born model and its various variants~\cite{hentschel_PoP_2023}, essentially leading to RPA-like behavior. Figure~\ref{fig:fsum_omegadomain} also suggests that for a given collision frequency model different dynamic response models will lead to different conclusions on physical quantities such as the Rayleigh weight $W_R$ or density.

Lastly, future work developing dielectric function models / dynamic structure factor models may wish to consider the following. Firstly, in practice, the CM model is not perfect. By including momentum conservation the closure issue in the continuity equation is corrected, but we have removed the dissipation that gives rise to a finite conductivity in a single component system~\cite{atwal_PRB_2002, Morawetz_PRE-momentum_2000}. Enforcing a different constraint on $\delta u$ may remove this trade-off; for example, constraining $\delta u$ to satisfy Onsager relations. 
Secondly, if both momentum conservation and a finite conductivity are desired then a multi-species model is needed~\cite{Morawetz_PRE-momentum_2000}. However, Garzo \textit{et al}.'s multi-species BGK equation~\cite{garzo_PhysFluidsA_1989} satisfies the Onsager relations, but the entropy does not uniformly increase~\cite{haack_JSP_2017} and Haack \textit{et al}.'s multi-species BGK equation~\cite{haack_JSP_2017, haack_PRE_2017} satisfies the conservation laws and the entropy increases, but it does not satisfy the Onsager relations~\cite{benilov_PRE_2024}. 
Thirdly, while the present work relies on the kinetic formalism other formalisms may be used. For example, the hydrodynamic formalism has been used elsewhere. Diaw and Murillo~\cite{Diaw_SciRep_2017} compute a single species electron DSF model from quantum hydrodynamic (QHD) equations for number and momentum conservation using dynamic density functional theory and recover a Dirac delta in the long wavelenth limit. Opposite this, Svensson et al.~\cite{Svensson_PRE_2025} compute a DSF from classic hydrodynamic equations for mass, momentum, energy conservation, introducing dissipation through ion-electron interaction and recover Mermin's Lorentz-like DSF in the long wavelength limit. 
Thus, the tradeoff between conservation laws and dissipation persists across both quantum and classical as well as hydrodynamic and kinetic formalisms, often manifesting in the long wavelength limit as a difference between Lorentz-like and Dirac-delta-like response. Future work ought to consider fundamental questions about the emergence of entropy from a conservative system in order to derive an appropriate dynamic response model for their multi-species system. 

\begin{acknowledgments}
Thomas Chuna acknowledges: Firstly, Hannah Bellenbaum and Pontus Svensson for literature references. Secondly, deceased MSU physicist F.D.C. Willard for lively discussions and insights into the foundational works of solid state physics. Thirdly, ChatGPT suggested the complex analysis techniques used to evaluate the second term in \eqref{eq:Mermin_fsum_app_helper2}.

This work has received funding from the European Union's Just Transition Fund (JTF) within the project \textit{R\"ontgenlaser-Optimierung der Laserfusion} (ROLF), contract number 5086999001, co-financed by the Saxon state government out of the State budget approved by the Saxon State Parliament. This work has received funding from the European Research Council (ERC) under the European Union’s Horizon 2022 research and innovation programme
(Grant agreement No. 101076233, "PREXTREME"). 
Views and opinions expressed are however those of the authors only and do not necessarily reflect those of the European Union or the European Research Council Executive Agency. Neither the European Union nor the granting authority can be held responsible for them. 
Tobias Dornheim gratefully acknowledges funding from the Deutsche Forschungsgemeinschaft (DFG) via project DO 2670/1-1.
Computations were performed on a Bull Cluster at the Center for Information Services and High-Performance Computing (ZIH) at Technische Universit\"at Dresden and at the Norddeutscher Verbund f\"ur Hoch- und H\"ochstleistungsrechnen (HLRN) under grant mvp00024.
\end{acknowledgments}


\appendix
\section{Explicitly evaluating the f-sum rule for Mermin's dielectric function in the long wavelength limit}
\label{app:complexanalysis}
In this appendix, we assume that the collision frequency is real and evaluate the f-sum rule~\eqref{eq:fsum_dielectric} for Mermin's model \eqref{eq:Merminlongwavelength} explicitly. Essentially, we evaluate the following expression,
\begin{align}\label{eq:Mermin_fsum_app}
    - \int_{-\infty}^{\infty} \frac{d\omega}{\omega_p} \,  \frac{\omega}{\omega_p}  \frac{  \nu \, \omega \, \omega_p^2}{ ( \omega^2 - \omega_p^2)^2 + \nu^2 \omega^2} \, ,
\end{align}
which is obtained by substituting Mermin's long wavelength expansion \eqref{eq:Merminlongwavelength} into the f-sum rule \eqref{eq:fsum_dielectric}. To evaluate \eqref{eq:Mermin_fsum_app}, we want to uncover the poles of the integrand by removing its dimensions and substituting variables. 

To remove dimension, we substitute $\nu = 1/\tau$ and normalize all the quantities by $\omega_p$, \textit{i.e.} $\tilde{\omega} = \omega / \omega_p$ and $\tilde{\nu} = \nu / \omega_p = (\tau \omega_p)^{-1}$. This yields 
\begin{align}
    - \int_{-\infty}^{\infty} d \tilde{\omega} \,  \frac{  \tilde{\omega}^2/ \tilde{\tau}}{ ( \tilde{\omega}^2 - 1)^2 +  (\tilde{\omega}/\tilde{\tau})^2} \, .
\end{align}
Now that all quantities are dimensionless, we will drop the tilde for the remainder of this appendix and reintroduce dimensionality at the end. Next, we multiply the numerator and denominator by $\tau^2$ and factor out $\omega^2$ from the denominator, which cancels the $\omega^2$ in the numerator and yields
\begin{align}
    \int_{-\infty}^{\infty} d \omega \,   \frac{-\tau }{ \tau^2 (\omega - \omega^{-1} )^2 + 1 } \, .
\end{align}
We use variable substitution $u = \tau(\omega - \omega^{-1})$ and $du = \tau (1 + \omega^{-2}) d\omega$, simplifying the expression to
\begin{align}
    - \int_{-\infty}^{\infty} d u \frac{\omega^2}{(\omega^2 + 1)}  \frac{1}{ u^2 + 1 } \, . \label{eq:Mermin_fsum_app_helper1}
\end{align}

Equation \eqref{eq:Mermin_fsum_app_helper1} contains the typical scaled Cauchy distribution modified by the $d\omega$ term, which has preserved an undesirable $\omega$ dependence. To evaluate this expression, we separate the integral into its parts by adding $0 = 1 -1$ to the numerator and recover,
\begin{align}
    &- \int_{-\infty}^{\infty} d u  \frac{1}{ u^2 + 1 } + \int_{-\infty}^{\infty} d u \frac{1}{(\omega^2 + 1)}  \frac{1}{ u^2 + 1 },
\end{align}
We make use of the indefinite integral
\begin{align}
    \arctan(u) = \int du \, (u^2 + 1)^{-1} \, ,
\end{align}
that is  commonly used to compute the Cauchy distribution's cumulative distribution function (CDF)~\cite{berg_book_2004}. We also undo the variable substitution to arrive at
\begin{align}
    - \left. \arctan\left(\tau \frac{\omega^2 - 1 }{\omega}\right) \right\vert_{-\infty}^{+\infty} + \int_{-\infty}^{\infty} d \omega \frac{\tau}{ \tau^2(\omega^2 - 1)^2 + \omega^2} \, , \label{eq:Mermin_fsum_app_helper2}
\end{align}

As desired, \eqref{eq:Mermin_fsum_app_helper2} exposes the poles of the integrand. The second term is a rational integral and can be computed using complex analysis techniques, therefore we factor the quadratic into its roots
\begin{subequations}
\begin{align}\label{eq:decomposed-integral}
    &- \left. \arctan\left(\tau \frac{\omega^2 - 1 }{\omega}\right) \right\vert_{-\infty}^{+\infty} + \int_{-\infty}^{\infty} d \omega \frac{\tau}{ \tau^2(\omega^2 - a)(\omega^2 - b)},
\end{align}
where 
\begin{align}
    \quad a = \frac{2 \tau^2 - 1 + \sqrt{1 - 4 \tau^2}}{2 \tau^2} \, ,
    \\b = \frac{2 \tau^2 - 1 - \sqrt{1 - 4 \tau^2}}{2 \tau^2} \, .
\end{align}
\end{subequations}
Two cases emerge for this expression $\tau=1/2$ (\textit{i.e.}, $a=b=-1$), $\tau \neq 1/2$ (\textit{i.e.}, $a\neq b$).

We consider the $\tau = 1/2$ case first, the expression for the f-sum rule is given.
\begin{align}
    &- \left. \arctan\left(\frac{\omega^2 - 1 }{2\omega}\right) \right\vert_{-\infty}^{+\infty} + \int_{-\infty}^{\infty} d \omega \frac{2}{(\omega^2 + 1)^2} \, ,
\end{align}
Simplify the first term by parameterizing $\omega$ such that $\tan \theta = \omega$ and $\theta = \arctan(\omega)$ and use the double angle trig identities, and anti-symmetry of $\arctan$ to produce, 
\begin{align}
\left. \left(\frac{\pi}{2} - 2 \arctan(\omega) \right)\right\vert_{-\infty}^{+\infty} + \int_{-\infty}^{\infty} d \omega \frac{2}{(\omega^2 + 1)^2} \, .
\end{align}
The $\pi$ term cancels with itself and the second term is a standard integral. Together this yields 
\begin{align}\label{eq:Merminfsumconvergence_t0p5_app}
    \left. \left( -2 \arctan(\omega)  + \arctan(\omega) + \frac{\omega}{\omega^2 +1}\right) \right\vert_{-\infty}^{+\infty} \, ,
\end{align}
which can be evaluated to give 
\begin{align}\label{eq:fsum_arctan_1}
     -\pi = -\arctan(+\infty) + \arctan(-\infty) \, .
\end{align}
Thus, for $\tau = 1/2$ the Mermin dielectric function satisfies the f-sum rule. 

Now we consider the $\tau \neq 1/2$ case for \eqref{eq:decomposed-integral}. We expand partial fractions to arrive at
\begin{align}
    &- \arctan\left( \tau \frac{\omega^2 - 1 }{\omega} \right) \nonumber
    \\ & \quad + \frac{\tau}{\tau^2(a-b)}\int_{-\infty}^{\infty} d \omega  \left(\frac{1}{ \omega^2 - a} -  \frac{1}{ \omega^2 - b} \right) \, .
\end{align}
Notice this expression is not valid at $\tau = 1/2$, since $a = b$. Next we again expand partial fractions to separate the roots of $\omega^2 -a$, (\textit{i.e.}, $\pm \sqrt{a}$) and use the elementary integral $\int d\omega \, 1/(\omega-\sqrt{a}) = \ln( | \omega - \sqrt{a} | )$ to arrive at,
\begin{align}
    &- \left.\arctan\left( \tau \frac{\omega^2 - 1 }{\omega} \right)\right\vert_{-\infty}^{+\infty}  \nonumber
    \\ & \quad + \left. \frac{\tau}{\sqrt{1 - 4\tau^2}} \left( \frac{1}{2 \sqrt{a}} \ln \left| \frac{\omega - \sqrt{a}}{\omega + \sqrt{a}}\right| \right) \right\vert_{-\infty}^{+\infty} \, , \nonumber
    \\ & \quad - \left. \frac{\tau}{\sqrt{1 - 4\tau^2}} \left( \frac{1}{2 \sqrt{b}} \ln \left| \frac{\omega - \sqrt{b}}{\omega + \sqrt{b}}\right| \right)  \right\vert_{-\infty}^{+\infty}\, , \label{eq:fsum_arctan_2}
\end{align}
where we have substituted ${a-b} = (\sqrt{1 - 4\tau^2}\,)/ \tau^2$. Evaluating \eqref{eq:fsum_arctan_2}, we recover that for $\tau \neq 1/2$ the Mermin dielectric function satisfies the f-sum rule. To see this, notice there is an absolute value surrounding the argument of the logarithm, which implies that it does not matter if $a$ and $b$ yield negative or complex values the large $\omega$ behavior scales as $\ln (1)$. This leaves only the $\arctan$ term, as is in \eqref{eq:fsum_arctan_1}. 

More rigorously, we Taylor expand the logarithmic terms in \eqref{eq:fsum_arctan_2} up to $\mathcal{O}[\omega^{-5}]$ to recover that,
\begin{align}
    \frac{\tau}{\sqrt{1 - 4\tau^2}} \left( \frac{1}{2 \sqrt{b}} \ln \left| \frac{\omega - \sqrt{b}}{\omega + \sqrt{b}}\right| \right)  & \approx \frac{\tau}{\sqrt{1-4\tau^2}}\frac{b - a}{3 \omega^3} \, , \nonumber
    \\ &=\frac{1}{\tau}\frac{1}{3 \omega^3} \,.
\end{align}
Compared to the Taylor expansion of $\arctan(\omega) \approx \pi/2 - 1/\omega$ the additional term is negligible. 

We have now resolved both cases and found that, as is expected of a function that scales like $\omega^{-2}$, both equation~\eqref{eq:Merminfsumconvergence_t0p5_app} and \eqref{eq:fsum_arctan_2} go as $\arctan$ at large $\omega$, this demonstrates the validity of \eqref{eq:Merminfsumconvergence}. To explicitly recover equation \eqref{eq:Merminfsumconvergence}, let $\tau \rightarrow 1/\nu$ and redimensionalize the $\arctan$ argument $(\omega^2 -1 ) / (\nu \omega) \rightarrow (\omega^2 - \omega_p^2 ) / (\nu \omega)$, which is obtained via $\omega \rightarrow \omega/ \omega_p$, and $\nu \rightarrow \nu/\omega_p$.

\bibliography{bibliography}

\end{document}